\documentstyle[11pt,aaspp4]{article}
\tighten
\newcommand\kms{km~s$^{-1}$}

\slugcomment{98/08/12 -- Accepted by AJ}

\begin{document}

\title{New \ion{H}{1}-detected Galaxies in the Zone of Avoidance}

\author{
L. Staveley-Smith,\altaffilmark{1}
S. Juraszek,\altaffilmark{2}
B.S. Koribalski,\altaffilmark{1}
R.D. Ekers,\altaffilmark{1} 
A.J. Green, \altaffilmark{2}
R.F. Haynes,\altaffilmark{1}
P.A. Henning,\altaffilmark{3}
M.J. Kesteven,\altaffilmark{1}
R.C. Kraan-Korteweg,\altaffilmark{4}
R.M. Price,\altaffilmark{1}
and
E.M. Sadler\altaffilmark{2}
}

\altaffiltext{1}{Australia Telescope National Facility, CSIRO, P.O.\ Box
76, Epping, NSW 2121, Australia}
\altaffiltext{2}{School of Physics, University of Sydney, NSW 2006,
Australia}
\altaffiltext{3}{Institute for Astrophysics, University of New Mexico,
800 Yale Blvd, NE, Albuquerque, NM 87131, USA}
\altaffiltext{4}{Departemento de Astronomia, Universidad de Guanajuato,
Apartado Postal 144, Guanajuato, Gto 36000, Mexico}

\begin{abstract}

  We present the first results of a blind \ion{H}{1} survey for
  galaxies in the southern Zone of Avoidance with a multibeam receiver
  on the Parkes telescope.  This survey is eventually expected to
  catalog several thousand galaxies within Galactic latitude
  $|b|<5\arcdeg$, mostly unrecognised before due to Galactic
  extinction and confusion.  We present here results of the first
  three detections to have been imaged with the Australia Telescope
  Compact Array (ATCA). The galaxies all lie near Galactic longitude
  325\arcdeg\ and were selected because of their large angular sizes,
  up to 1\fdg3. Linear sizes range from 53 to 108 kpc.  The first
  galaxy is a massive $5.7\times 10^{11} {\rm M}_{\sun}$ disk galaxy
  with a faint optical counterpart, SGC 1511.1--5249. The second is
  probably an interacting group of galaxies straddling the Galactic
  equator. No optical identification is possible.  The third object
  appears to be an interacting pair of low column density galaxies,
  possibly belonging to an extended Circinus or Centaurus A galaxy
  group. No optical counterpart has been seen despite the predicted
  extinction ($A_B=2.7$ -- $4.4$ mag) not being excessive.  We discuss
  the implications of the results, in particular the low \ion{H}{1}
  column densities ($\sim 10^{19}~{\rm atoms\ cm}^{-2}$) found for two
  of the three galaxies.

\end{abstract}


\section{INTRODUCTION}

Within a Galactic latitude of $\pm 20\arcdeg$, extinction and
foreground confusion can considerably affect the quality of optical
studies of galaxies.  Within $\pm 10\arcdeg$, most optical observations
are challenging, and observations at infrared wavelengths become
competitive with optical observations for the purposes of galaxy
identification (e.g.\ Saunders et al. 1994). Within $\pm 5\arcdeg$ of
the Plane, even infrared observations are impossible for most
extragalactic studies.  However, in this latitude range, \ion{H}{1}
observations at 21 cm wavelength have been used with considerable
success, since the Galactic neutral hydrogen has optical
depth only over the narrow velocity range covered by its emission.  Blind
searches in the Galactic Plane, or Zone of Avoidance (ZOA), were
pioneered by Kerr \& Henning (1987) and a survey with the Dwingeloo 25 m
telescope has been successful in discovering several new nearby
galaxies, including Dwingeloo-1 (Kraan-Korteweg et al. 1994; Henning
et al. 1998).

In the past, large-area blind \ion{H}{1} surveys have had limited
sensitivity because of the time it takes to scan significant areas of
sky.  However, the new Parkes 21 cm multibeam receiver with its array
of 13 feed horns (Staveley-Smith et al. 1996) allows much more rapid
mapping of the nearby Universe. We have begun a survey with
this receiver for galaxies within $\pm5\arcdeg$ of the southern
Galactic Plane.  This survey should eventually produce a list of
several thousand new galaxies behind the Plane, and give us
information about massive nearby galaxies such as Dwingeloo-1
(Kraan-Korteweg et al. 1994) and Circinus (Freeman et al. 1977) which
may have been previously overlooked. It will also help fill in our
knowledge of the structure of the Centaurus-Pavo and Abell 3627
superclusters (Fairall 1988, Kraan-Korteweg et al.\ 1996) and the
Puppis and Vela regions (Kraan-Korteweg \& Huchtmeier 1992, Lahav et
al.\ 1993, Kraan-Korteweg \& Woudt 1994, Yamada et al.\ 1994) as they
disappear behind the Galaxy, as well as uncover unknown large-scale
structure across or behind the Galactic Plane.

In this paper, we present results for three galaxies found during the
course of the survey. These galaxies are selected from a preliminary list of
$\sim 50$ galaxies found near longitude 325\arcdeg\ (Juraszek et al. 1998).
Each of the galaxies is
resolved in the $14'$ beam of the Parkes telescope, and thus must be
fairly nearby. In \S~2, we describe the Parkes survey observations for the 
three  galaxies. In \S~3, we
present the more detailed ATCA observations and optical CCD observations.
Finally, in \S~4, we discuss the implications of the 
findings.

\section{MULTIBEAM SURVEY OBSERVATIONS}

The ZOA survey began at the Parkes telescope\footnote{The Parkes 
Telescope and the Australia Telescope are funded by the Commonwealth of
Australia for operation as a National Facility managed by CSIRO.}
 on 1997 March 22, and is expected to be completed by early
1999. The survey will map the southern ZOA between longitudes
$212^{\circ}$ and $36^{\circ}$, and between latitudes 
$-5^{\circ}$ and $+5^{\circ}$. Data are taken by scanning the
telescope in Galactic longitude. Each scan is of length
8\arcdeg\ with the position angle of the multibeam receiver lying
nominally at 15\arcdeg\ from the Galactic equator. The final survey will
be heavily oversampled in the spatial domain.

\begin{table}[htbp]
  \begin{center}
    \leavevmode
     \begin{tabular}{lccc}
\hline\hline\\
Property   & J$1514-52$ & J$1532-56$ & J$1616-55$ \\[2mm]
\hline\\
RA (J2000)$^{\dagger}$ & $15^{\rm h}14^{\rm m}34^{\rm s}$ & 
               $15^{\rm h}32^{\rm m}43^{\rm s}$ &
               $16^{\rm h}16^{\rm m}49^{\rm s}$ \\
DEC (J2000)$^{\dagger}$& $-52\arcdeg 59\arcmin 25\arcsec$ &
               $-56\arcdeg 07\arcmin 30\arcsec$ &
               $-55\arcdeg 44\arcmin 57\arcsec$ \\
Galactic longitude $^{\dagger}$  & 323\fdg59 & 324\fdg07 & 329\fdg06 \\
Galactic latitude  $^{\dagger}$  & $4\fdg04$ & $-0\fdg02$ & $-3\fdg67$ \\
Heliocentric velocity, $cz$ (\kms) $^*$ & 1452 & 1383 & 421 \\
Velocity width at 50\% (\kms) $^*$ &  441 &   90 &  55 \\
Distance ($H_{\circ}=75\ {\rm km\ s}^{-1} {\rm Mpc}^{-1}$) (Mpc)
               & 17.0 & 16.0 & 3.7 \\
\ion{H}{1} Diameter (kpc)   $^{\dagger}$           & 53  & 108  & 86 \\
Peak \ion{H}{1} column density (atoms cm$^{-2}$) $^*$ & $16.3\times 10^{19}$ &
                                                      $2.0\times 10^{19}$  &
                                                      $0.8\times 10^{19}$  \\
\ion{H}{1} mass ($M_{\sun}$) $^*$ & $8.4\times 10^9$ & 
                                  $2.1\times 10^9$ &
                                  $8.9\times 10^7$ \\
Inclination  $^{\dagger}$ & 62\arcdeg & -- & -- \\
Position Angle $^{\dagger}$ & 156\arcdeg & -- & -- \\
Absorption, $A_B$(dust) & 4.3 mag & 57 mag & 2.7 mag \\
Absorption, $A_B$(gas) &  3.4 mag & 16 mag & 4.4 mag \\
Type         & SBb? & Interacting? & Interacting LSB? \\
& \\
\hline

     \end{tabular}
    \caption{The three large \ion{H}{1}-detected galaxies with multibeam 
      and ATCA data. Parameters measured from the multibeam data are
      indicated with an asterisk (*); those from the follow-up ATCA
      data with a dagger ($\dagger$).  Velocity widths are measured at
      the 50\% level from spatially integrated and Hanning-smoothed
      spectra (26 \kms\ resolution).  \ion{H}{1} diameters are
      measured at 1 M$_{\odot}$ pc$^{-2}$ ($1.26\times 10^{20}$
      cm$^{-2}$) for J$1514-52$, and at 10\% of the peak ATCA column
      density for J$1532-56$ and J$1616-55$ ($1.0\times 10^{19}$
      cm$^{-2}$, and $2.1\times 10^{18}$ cm$^{-2}$, respectively). The
      optical absorption estimates are based on the data of Schlegel
      et al.\ (1998) (dust) and Dickey \& Lockman (1990) (gas).}
    \label{t:newgal}
  \end{center}
\end{table}

\begin{figure}
\epsscale{0.45}
\plottwo{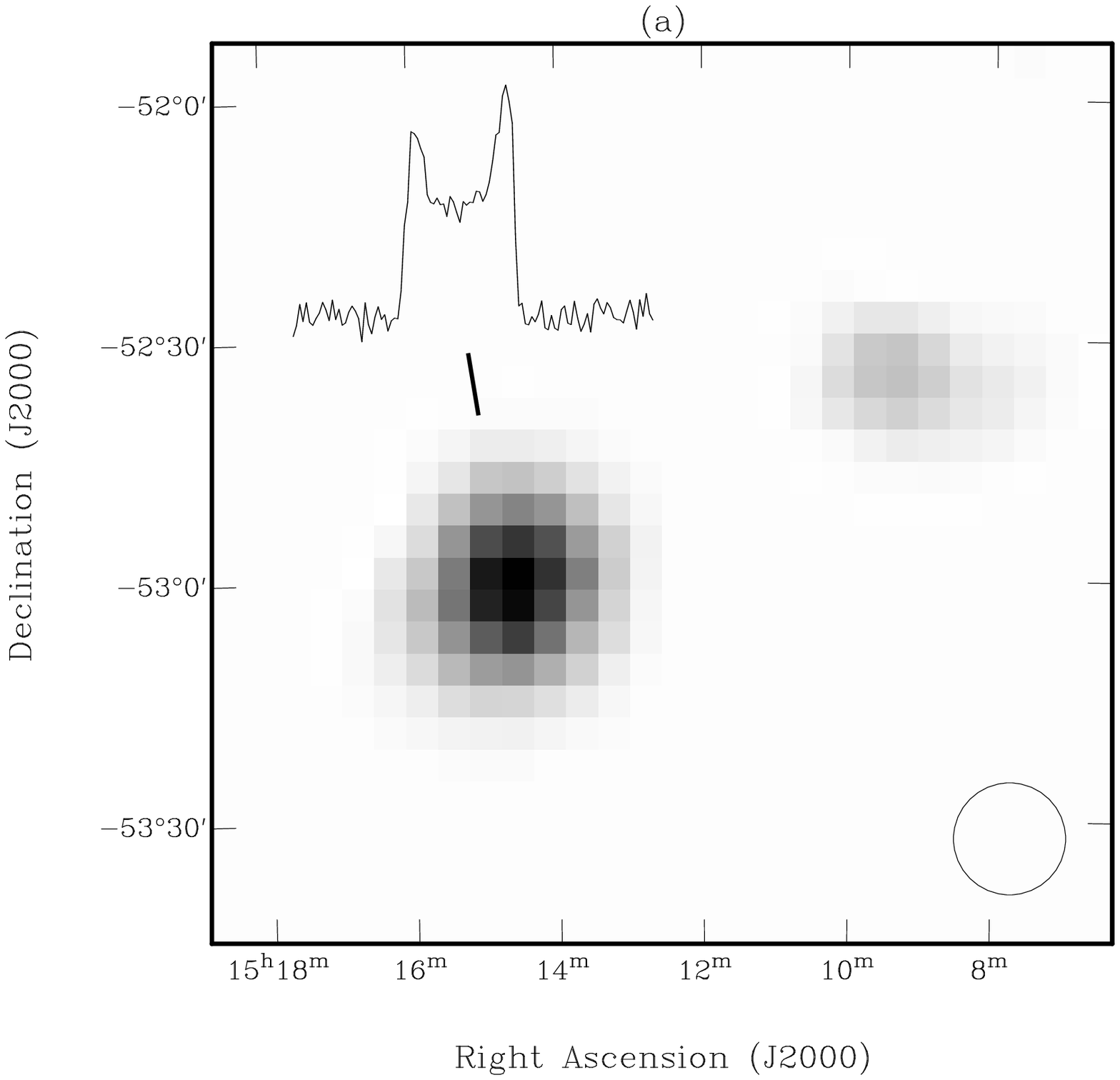}{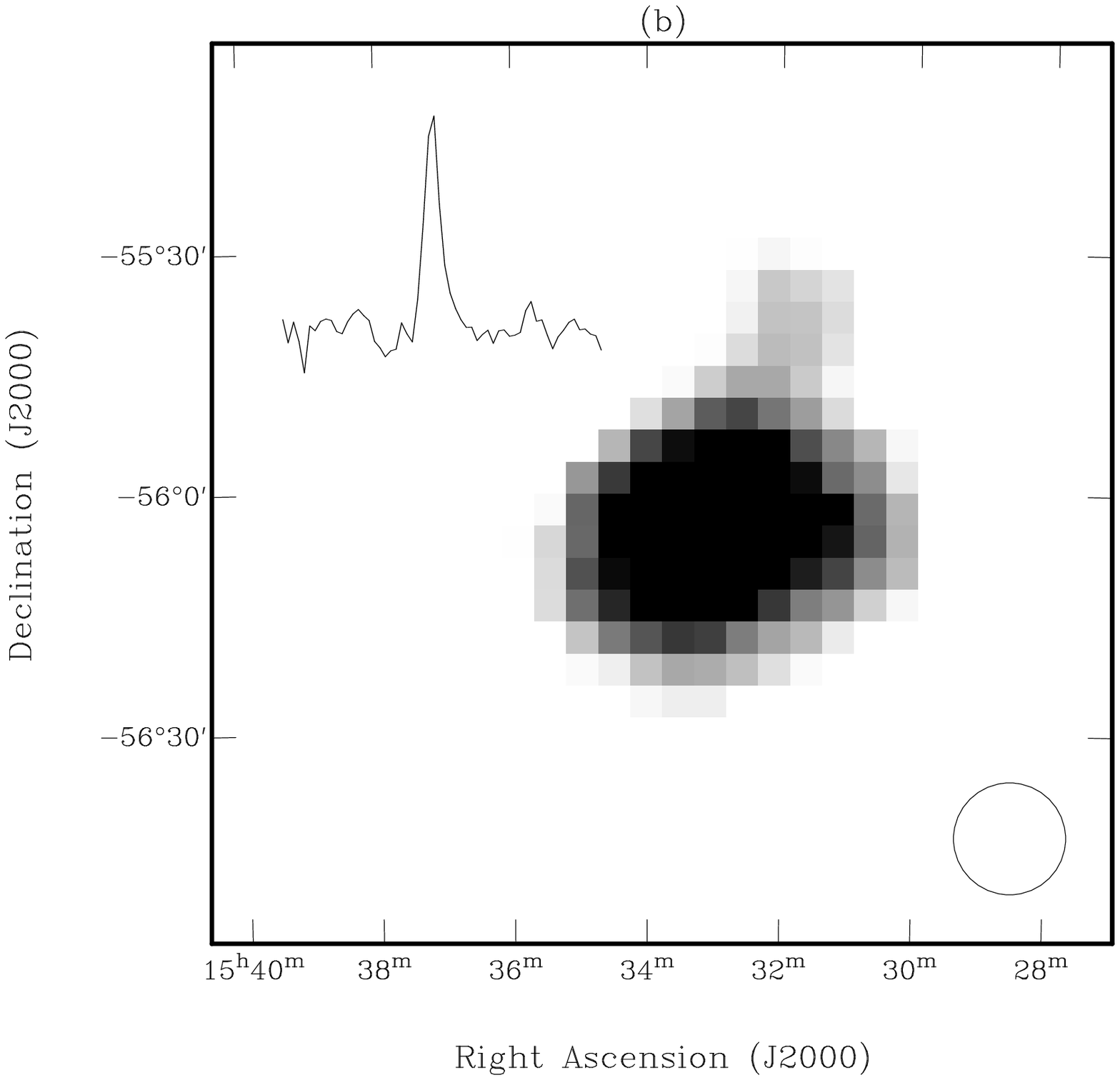}
\epsscale{0.45}
\plotone{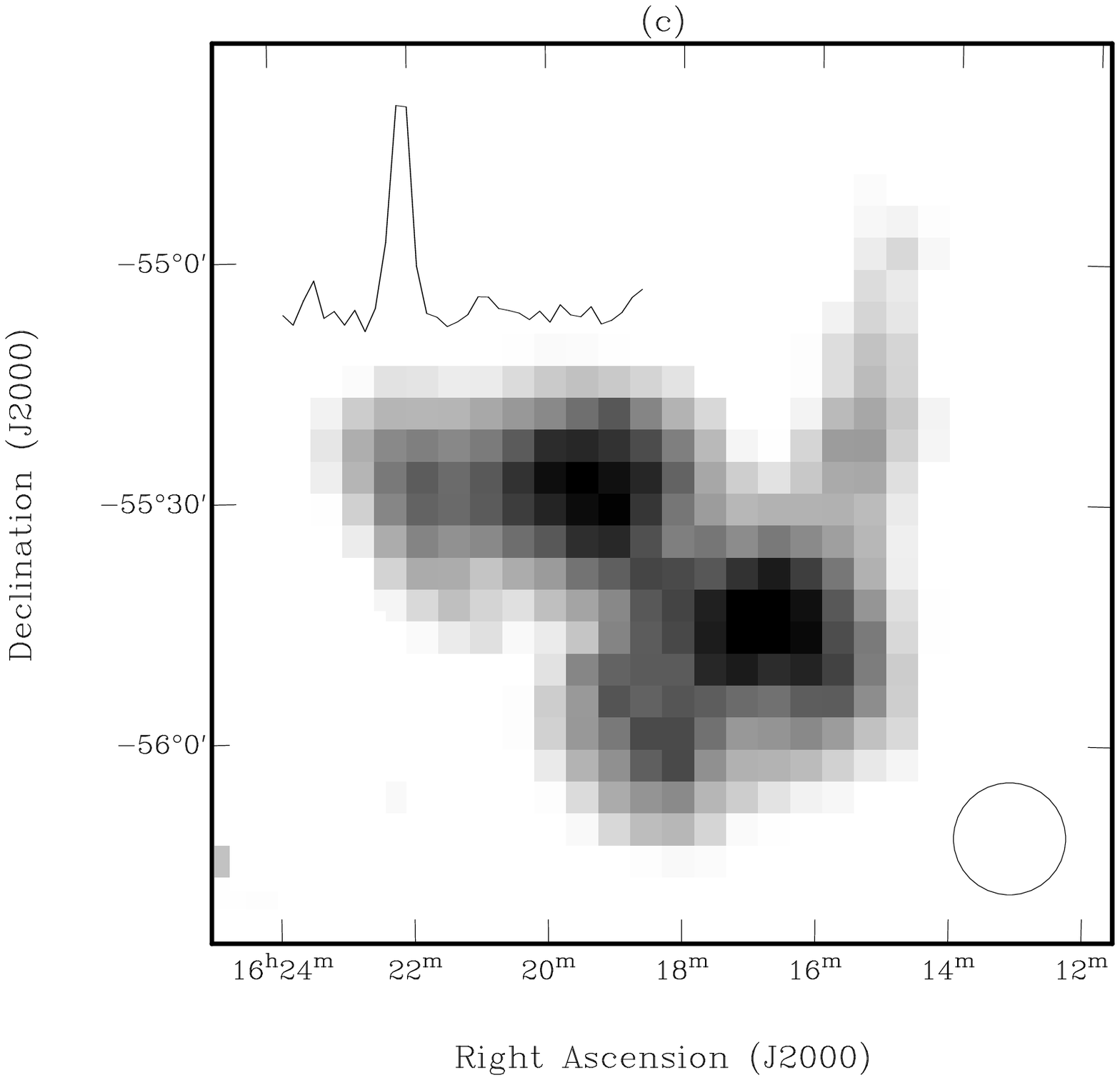}
\caption{Parkes multibeam column density images, with velocity contours
  overlayed, for the three galaxies: (a) J$1514-52$ (note the companion,
  ESO 223-G012 to the northwest); (b) J$1532-56$, and (c) J$1616-55$. The
  field size is 1\fdg9 in each case. The integrated velocity profiles for the
  objects is overlayed at the top left of each panel, and the Parkes
  beam is drawn in the lower right. The peak column densities range
  from $8\times 10^{18}$ to $2\times 10^{20}$ atoms cm$^{-2}$ (see
  Table~\ref{t:newgal}).}
\label{f:mb-moment}
\end{figure}

The multibeam correlator has a bandwidth of 64 MHz and covers the
frequency range 1362.5 to 1426.5 MHz. This corresponds to a velocity
($cz$) range of approximately $-1200$ to $12700$ \kms. The channel
spacing is 13 \kms, but to suppress ringing caused by the strong
narrow-band Galactic \ion{H}{1} emission, the data shown here are
Hanning smoothed to a resolution of 26 \kms.  The average FWHP
beamwidth of the multibeam receiver is $14'$, though the gridding
process increases this to $17'$. The average system temperature is 20
K. The data used here have an rms noise of about 15 mJy beam$^{-1}$,
corresponding to an \ion{H}{1} column density of around
$7\times 10^{17}$ atoms cm$^{-2}$ in each 26 \kms\ resolution element.
The data were calibrated and baseline-subtracted using a real-time
system based on aips++ (Barnes et al.\ 1997). Cubes were also made
using specially developed aips++ routines (Barnes 1997). Miriad, aips
and aips++ were used for cube concatenation, Hanning smoothing and
making moment maps. All spectral calibration and cube-forming routines
use robust statistics. In combination with high oversampling of the
sky, this allows enormous suppression of any non-continuous
interference.

The galaxies chosen for follow-up observations are shown in
Table~\ref{t:newgal} together with positions and other measured
properties. The column-density images for J$1514-52$, J$1532-56$ and
J$1616-55$ from the multibeam survey are shown in
Figure~\ref{f:mb-moment}. All the galaxies are near $\ell=325\arcdeg$
and are well-resolved, implying large overall sizes.  Only J$1514-52$,
at $b=4\fdg04$ has a previously identified optical counterpart, SGC
$1511.1-5249$ (Corwin et al.  1985).  J$1514-52$ is also notable for
having a close companion, ESO~223--G~12, visible in the multibeam
image (Figure~\ref{f:mb-moment}). The companion has an almost
identical velocity and velocity width to J$1514-52$.  The \ion{H}{1}
masses of the three new galaxies range from $9\times 10^7 {\rm
  M}_{\sun}$ (J$1616-55$) to $8\times 10^9 {\rm M}_{\sun}$
(J$1514-52$) and peak column densities from $8\times 10^{18}\ {\rm
  atoms\ cm}^{-2}$ (J$1616-55$) to $2\times 10^{20}\ {\rm atoms\ 
  cm}^{-2}$ (J$1514-52$) (at the multibeam resolution). The linewidths
range from 55 \kms\ (J$1616-55$) to 441 \kms\ (J$1514-52$). The large
\ion{H}{1} mass and linewidth for J$1514-52$ are typical of a massive
spiral galaxy. The low \ion{H}{1} mass and linewidth for J$1616-55$
are typical of a low-mass dwarf system. However, the large linear size
(86 kpc, see below) and low column density of J$1616-55$ is not
typical of an \ion{H}{1}-rich dwarf galaxy, a point to which we return
in \S~4.

\section{RADIO AND OPTICAL FOLLOW-UP OBSERVATIONS}

\begin{table}[htbp]
  \begin{center}
    \leavevmode
     \begin{tabular}{lccc}
\hline\hline\\
     & J$1514-52$ & J$1532-56$ & J$1616-55$ \\[2mm]
\hline\\
Observation Date(s) & 97/06/08 & 97/06/08, 97/10/01 & 97/06/06, 97/07/16 \\
Array(s) & 750A & 750A, 375 & 750A, 122B \\
Integration time (h) & 4 & 4.5+9 & 13+12 \\
Bandwidth (MHz) & 8 & 8 & 8 \\
Number of pointings  & 1 & 1 & 12 \\
Velocity resolution (\kms) & 6.6 & 6.6 & 6.6/26.4 \\
Synthesised beamwidth &  $1\farcm4\times0\farcm8$ & $2\farcm2\times2\farcm2$ &
                         $4\farcm5\times4\farcm5$ \\
Position angle & 31\arcdeg & 0\arcdeg & 0\arcdeg \\
Combined with multibeam data? & $\times$   & $\times$  & $\surd$ \\
& \\
\hline

     \end{tabular}
    \caption{ATCA journal of observations.}
    \label{t:journal}
  \end{center}
\end{table}

High-resolution \ion{H}{1} and optical CCD observations were 
made of the multibeam-detected galaxies. Table~2 lists the observing 
parameters for the ATCA \ion{H}{1} observations. The CCD observations are
discussed in the text.

\subsection{J$1514-52$}

\begin{figure}[t]
\epsscale{0.6}
\plotone{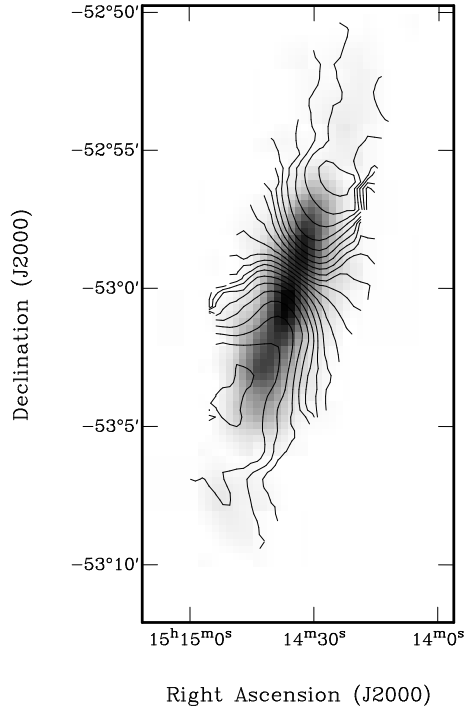}
\caption{ATCA column density map of J$1514-52$ with contours of equal velocity
overlayed. The peak brightness is 6.81 Jy  beam$^{-1}$ km s$^{-1}$,
corresponding to a column density of $1.8\times 10^{21}$ atoms cm$^{-2}$
(higher than the multibeam value, due to higher spatial resolution).
The velocity contours are in intervals of 20 km s$^{-1}$
starting from 1240 km s$^{-1}$ in the northwest and finishing at 
1640 km s$^{-1}$ in the southeast. The beam size is 
1\farcm4 $\times$ 0\farcm8 (natural weighting).}
\label{f:atca-1514-moment}
\end{figure}

\begin{figure}[t]
\epsscale{0.8}
\plotone{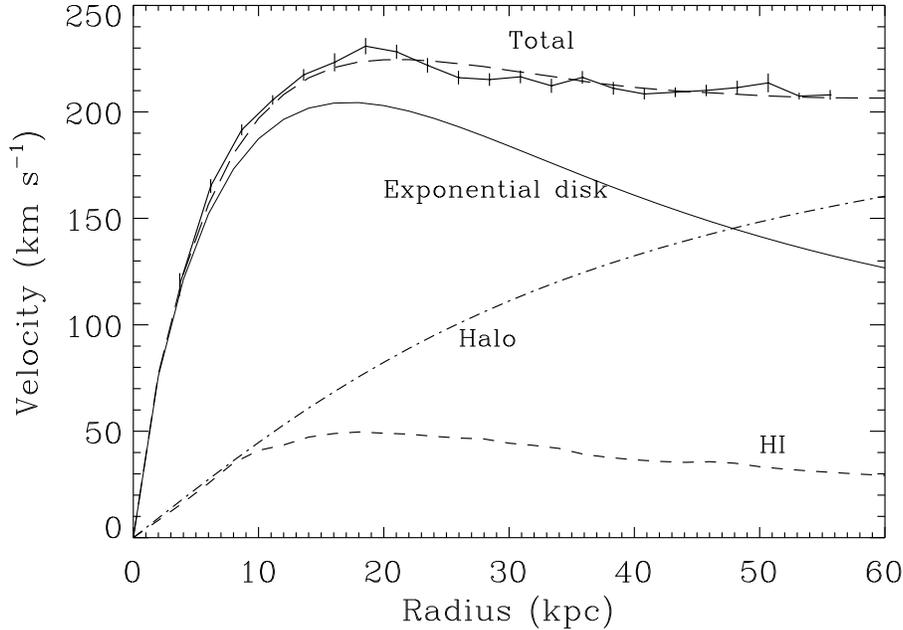}
\caption{Observed rotation curve for J$1514-52$ (thick solid line with error
bars) compared with a model (long dashed line) which comprises: an
exponential disk of scale length 8 kpc (thin solid line), an \ion{H}{1}
+He (mass ratio 1:0.3) disk (short dashed line), and a pseudo-isothermal
dark halo of core radius 30 kpc. The total mass of all
components within a radius of 60 kpc is $5.7\times 10^{11}$ M$_{\odot}$.}
\label{f:rotcur}
\end{figure}

The ATCA column density image is shown in
Figure~\ref{f:atca-1514-moment}.  J$1514-52$ is clearly resolved and
is a highly inclined, rotating disk galaxy.  The \ion{H}{1} mass
measured from the primary-beam-corrected ATCA column density image is
$4.8\times 10^9$ M$_{\odot}$. This is only 60\% of the \ion{H}{1} mass
measured in the multibeam observations ($8.4\times 10^9$ M$_{\odot}$,
Table~1).  The difference is due to the lack of short spacings in the
750A array, and primary beam attenuation\footnote{The FWHP primary
  beamwidth of the ATCA at 21~cm is 34\arcmin.} which forces the
signal below the threshhold level for detection of significant
\ion{H}{1}.  The diameter at a projected \ion{H}{1} surface density of
1 M$_{\odot}$ pc$^{-2}$ ($1.26\times 10^{20}$ cm$^{-2}$) is 53 kpc
(Table~1). However, the total \ion{H}{1} extent is much larger. A
tilted ring analysis (which gives the rotation curve shown in
Figure~\ref{f:rotcur}) extends to a radius of 55 kpc.  Given the
relatively short observation and the fact that the ATCA primary beam
correction is significant, the galaxy may extend considerably further still.
The kinematically derived inclination is 62\arcdeg\ and the pa is
156\arcdeg.  The dynamical center is given in Table~\ref{t:newgal},
and is only 6\arcsec\ from the precise optical position listed by
Woudt \& Kraan-Korteweg (1998).

An approximate decomposition of the rotation curve into an exponential
disk, \ion{H}{1} disk, and dark halo is reasonably successful
(Figure~\ref{f:rotcur}). The portion of the rotation curve due to
\ion{H}{1} was fixed according to the measured ATCA \ion{H}{1} density
profile (corrected for primary beam attenuation) but normalised so
that the \ion{H}{1} mass agrees with the multibeam mass in Table~1. It
is assumed that the \ion{H}{1} lies in a flat, infinitely thin disk. A
30\% (by mass) He contribution was also allowed for. The steep initial
rise in the rotation curve was well-fitted by an infinitely thin
exponential disk with a scale length of 8 kpc and central density 500
M$_{\odot}$ pc$^{-2}$ (and total mass $2.0\times 10^{11}$
M$_{\odot}$). Finally, a pseudo-isothermal halo (e.g.\ Walsh et al.
1997) with a core radius of 30 kpc and central density of $10^{-3}$
M$_{\odot}$ pc$^{-3}$ was required to fit the outer part of the
rotation curve.

J$1514-52$ is therefore a very massive system. The mass of the dark halo
within a radius of 60 kpc is $3.6\times 10^{11}$ M$_{\odot}$. The combined
mass of the halo and both disk components is $5.7\times 10^{11}$ M$_{\odot}$.
The gentle decline of rotation velocity even at the largest
radii suggests that the extent and total mass of the galaxy are
significantly larger.

\begin{figure}
\epsscale{0.8}
\plotone{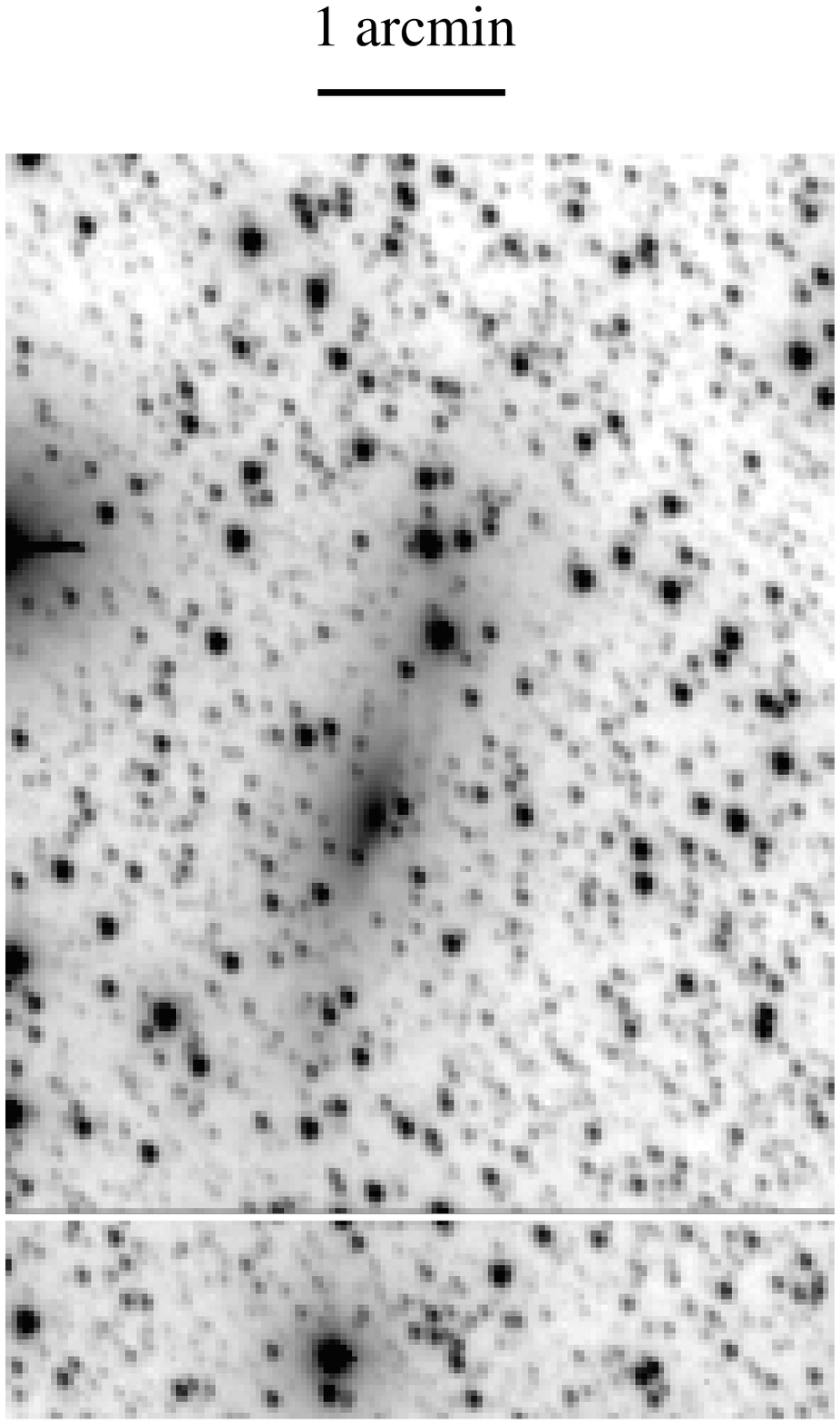}
\caption{$I$-band image of J$1514-52$ taken with the MSSSO 40 inch
telescope. North is up, east to the left.}
\label{f:1514_mssso}
\end{figure}

Optically, this galaxy is quite visible on sky-survey plates despite
its Galactic latitude of 4\fdg04. An $I$-band image taken with the
Mount Stromlo and Siding Spring Observatories (MSSSO) 40-inch
telescope on 1997 July 4 is shown in Figure~\ref{f:1514_mssso}.  The
exposure time is 16 min, and the pixel size (after $3\times 3$
binning) is 1\farcs8. The maximum optical diameter of J$1514-52$ on
Figure~\ref{f:1514_mssso} is 220\arcsec\ (17 kpc).  Woudt \&
Kraan-Korteweg (1998) list this galaxy as WKK 4748 in their catalog
and give a diameter of $212\arcsec \times 56 \arcsec$ and a magnitude
of $B_J=13.9\pm0.5$ mag. The Galactic column density of $4.6\times
10^{21}$ atoms cm$^{-2}$ (Dickey \& Lockman 1990) gives rise to an
estimated absorption $A_B=3.4$ mag. More recent reddening estimates
based on modelling of IRAS and DIRBE far-infrared data (Schlegel et al.
1998) suggests $A_B=4.3$ mag
(Table~\ref{t:newgal}) and therefore an absolute magnitude
$M_B\approx -21.6$ mag.  The corresponding \ion{H}{1} mass-to-light
ratio $M_H/L_B \approx 0.13$ is below the median value of 0.20 --
0.21 for the Sab/Sb samples listed by Roberts \& Haynes (1994), but well
within the range of these galaxies. The
rotation curve analysis above also gives a reasonable value for the
disk mass-to-light ratio $M_{\rm disk}/L_B \approx 3.1$.

With its optical size, J$1514-52$ should have been be included in the
Lauberts (1982) ESO catalog. However, it seems to have first been
listed as a galaxy by Corwin et al. (1985) who catalog it as SCG
1511.1-5249.  Improved coordinates are given by Spellman et al.
(1989).  It is a very strong infrared source (IRAS 15109-5248) with a
$100 \mu$m flux density of 60 Jy. Its far-infrared luminosity of
$1.0\times 10^{10} L_{\odot}$ lies in between two other moderately
active galaxies: Circinus ($5.7\times 10^9 L_{\odot}$) and NGC 253
($1.3\times 10^{10} L_{\odot}$).  The far-infrared luminosity may be
indicative of nuclear activity, possibly triggered by the nearby
companion, ESO 223-G012, which also appears in
Figure~\ref{f:mb-moment}.  J$1514-52$ has also been listed as a galaxy
candidate by Yamada et al (1993), Weinberger et al. (1995) and Takata
et al. (1996).

\subsection{J$1532-56$}
\label{s:1532-56}

\begin{figure}[t]
\plottwo{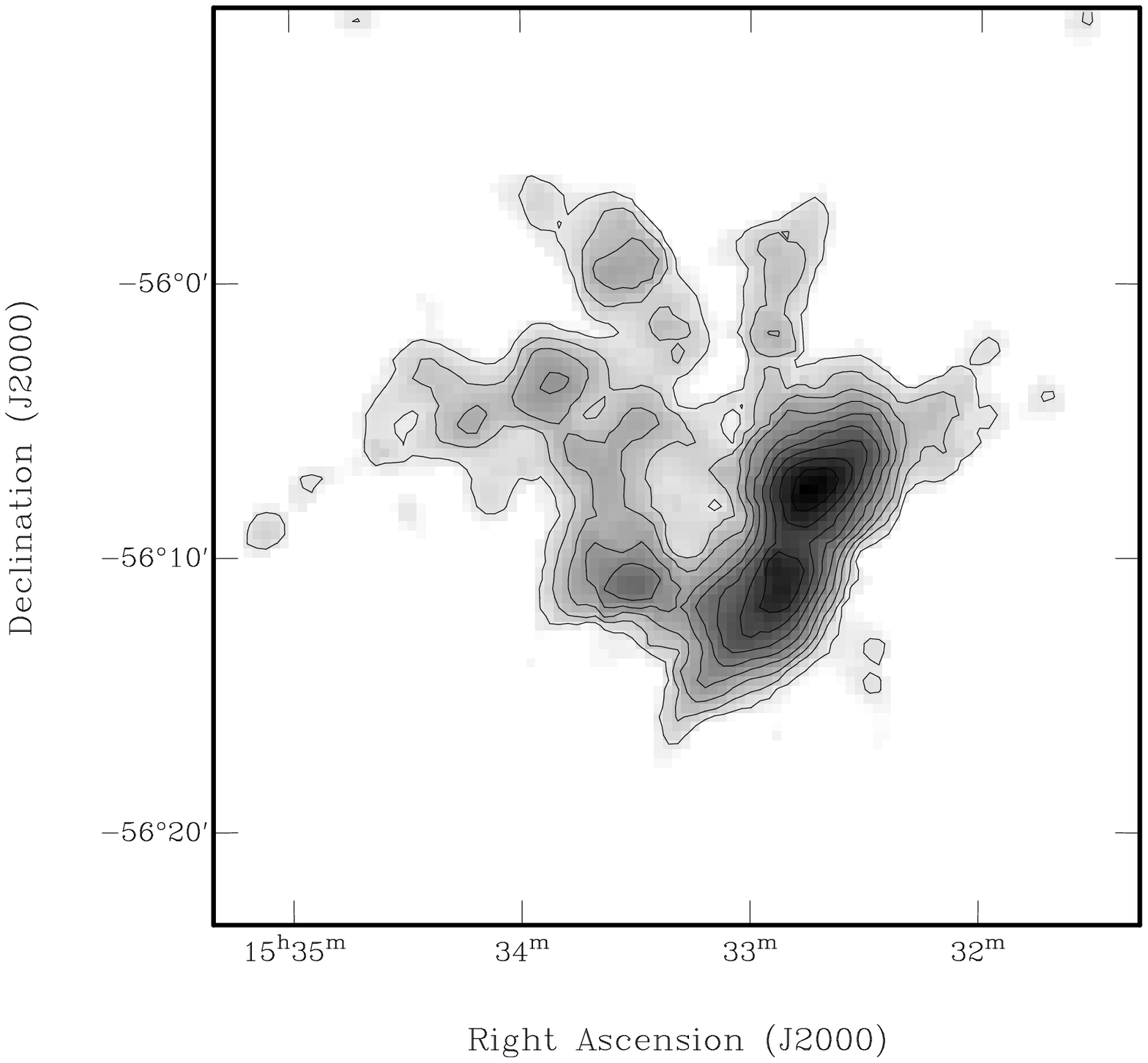}{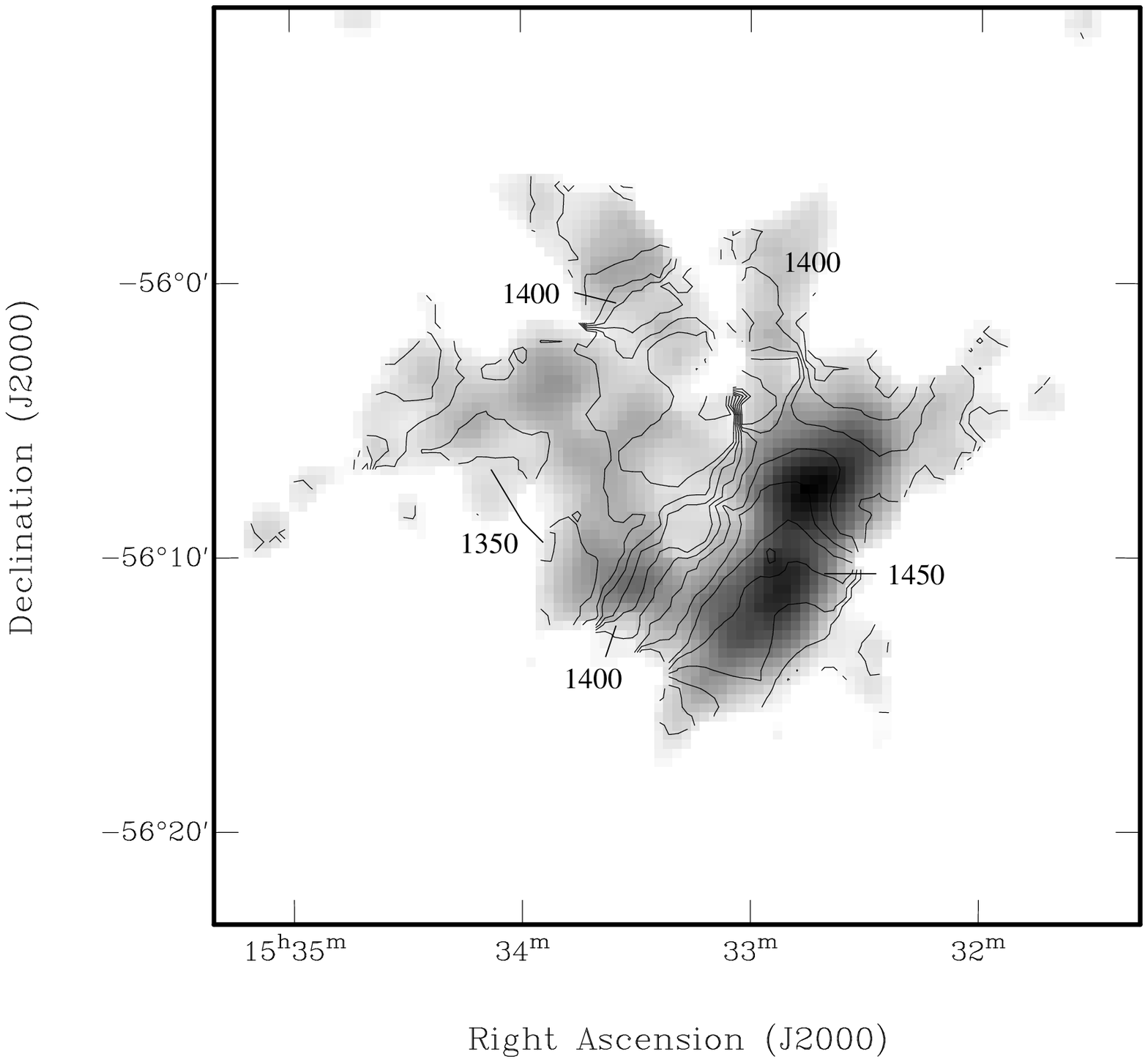}
\caption{Column density images of J$1532-56$ with contours of \ion{H}{1} 
column density (left) and velocity (right) overlayed. 
The column density contours are in interval of 0.16 Jy beam$^{-1}$ km s$^{-1}$
starting from 0.16 Jy beam$^{-1}$ km s$^{-1}$ ($10^{19}$ atoms cm$^{-2}$). 
The peak column density is $1.0\times 10^{20}$ atoms cm$^{-2}$. The velocity
contours are spaced by 10 km s$^{-1}$. The highest velocities occur in the 
south (1480 \kms) and north (1440 \kms); the lowest velocities in the east
(1360 \kms). Contour values are marked at 50 \kms\ intervals.
The ATCA beam is 2\farcm2.}
\label{f:atca-1532-moment}
\end{figure}

J$1532-56$ required more ATCA integration time and a shorter array to
detect (see Table~2).  The column-density image is shown in
Figure~\ref{f:atca-1532-moment}.
In \ion{H}{1}, J$1532-56$ is resolved into a relatively bright region
at RA $15^{\rm h}32^{\rm m}43^{\rm s}$, DEC $-56\arcdeg 07\arcmin
30\arcsec$ (J2000) which is surrounded by a diffuse \ion{H}{1} halo
and a bright extension of length $\sim 6\arcmin$ (28 kpc) at pa
150\arcdeg.  At the southern end of the extension, an arm-like feature
twists back in a manner reminiscent of the tidal tails
in the M81 group of galaxies (Yun et al.\ 1994). In the ATCA image,
the overall system appears to have a diameter of 108 kpc at $10^{19}$
atoms cm$^{-2}$, which is very similar to the 105 kpc diameter of the
M81/M82/NGC3077 system as measured by the VLA (Yun et al. 1994). The
extent of J$1532-56$ is probably underestimated by the ATCA
observations because of the primary beam cutoff. The multibeam data
suggest a value as high as 200 kpc.  The ATCA peak column
density is $10^{20}$ atoms cm$^{-2}$ and the average column density in
the tail is $3\times 10^{19}$ atoms cm$^{-2}$.  Both these values are
around two orders of magnitude smaller than for the M81 group, though
some of this is may be due to the large resolution difference.

\begin{figure}[t]
\epsscale{0.6}
\plotone{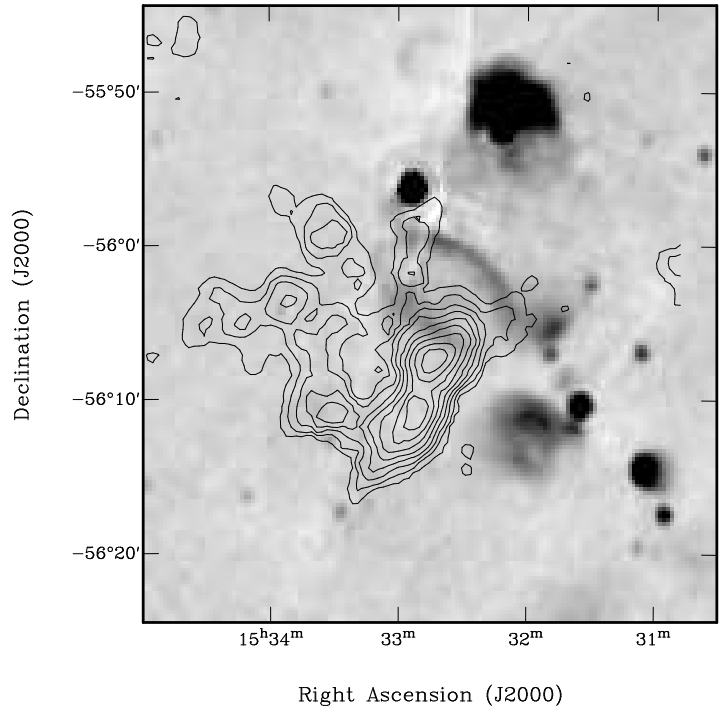}
\caption{An 843 MHz MOST image (Green et al. 1998) of the region around 
  J$1532-56$, with the ATCA \ion{H}{1} column-density map superimposed
  (contours as in Fig.~\ref{f:atca-1532-moment}). The galaxy lies
  close to a probable SNR, MSC 324.1+0.1 (the elliptical ring)
  (Whiteoak \& Green 1996), and to the \ion{H}{2} regions
  G324.192+0.109 (the bright source north of J$1532-56$) and
  G324.147+0.231 (the extended object at Dec $-55\arcdeg 50\arcmin)$.
  The other extended source at RA $15^{\rm h}32^{\rm m}$, Dec
  $-56\arcdeg 12\arcmin$ is unidentified.}
\label{f:1533-atca-most}
\end{figure}

J$1532-56$ lies at $(\ell, b) = (324\fdg07, -0\fdg02)$ where the estimated
absorption is $A_B=16$ -- 57 mag (Table~\ref{t:newgal}), thus making
optical and IR detection unlikely.  An 843 MHz radio continuum
image from the Molonglo Observatory Synthesis Telescope (MOST, Green
et al. 1998) in Figure~\ref{f:1533-atca-most} reveals a close
proximity (separation 15\arcmin) to the extended \ion{H}{2} region
G324.147+0.231, and even closer to the compact \ion{H}{2} region
G324.192+0.109 (Caswell \& Haynes 1987). This and the fact that the
heliocentric velocity of 1380 \kms\ corresponds to a frequency (1413.9
MHz) which is reasonably close to the H263$\delta$ recombination line
rest frequency (1413.6 MHz) prompted us to check whether were being
confused by Galactic emission. However, within the 64 MHz bandwidth of
the multibeam data, we failed to find any other recombination lines.
The absence of sufficient continuum emission at the position of
J$1532-56$ also makes this explanation untenable.

More interestingly, Figure~\ref{f:1533-atca-most} also shows J$1532-56$
to be close to an elliptical ring listed as a Galactic supernova
remnant (SNR) candidate by Whiteoak \& Green (1996). However, we think
it is unlikely that J$1532-56$ is a high-velocity cloud (HVC) associated with
an expanding SNR. The velocity of J$1532-56$ in the Galactic rest frame is
1251 \kms, which is well outside the range, $\pm 314$ \kms, quoted by
Stark et al.\ (1992) for Galactic HVCs. A search for molecular gas may 
help to evaluate this possibility.

\subsection{J$1616-55$}

\begin{figure}[t]
 \epsscale{0.6}
 \plotone{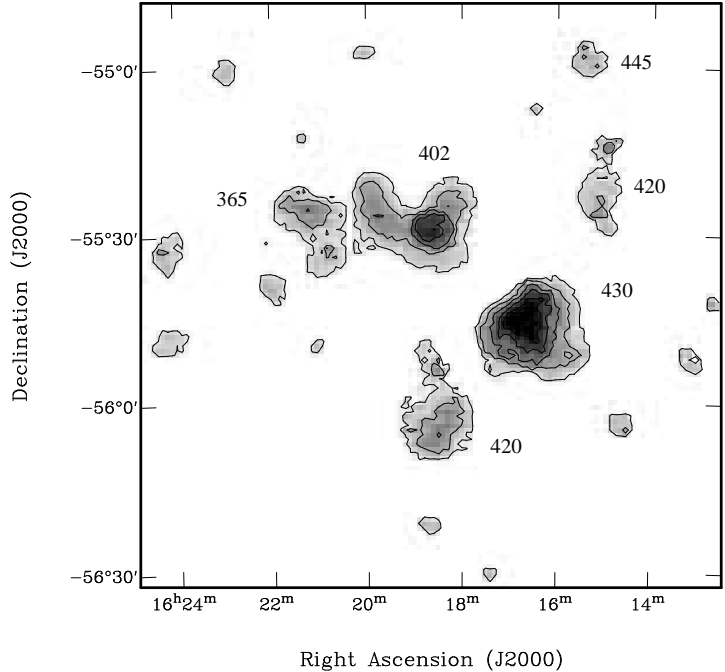}
\caption{The ATCA column density image for J$1616-55$ (natural weighting) 
  with contours at intervals of 0.2 Jy beam$^{-1}$ km s$^{-1}$,
  starting at 0.2 Jy beam$^{-1}$ km s$^{-1}$ ($3\times 10^{18}$ atoms
  cm$^{-2}$). The peak column density is $2.1\times 10^{19}$ atoms
  cm$^{-2}$ with the restoring beam used of 4\farcm5. This image is a
  12 point mosaic with the ATCA and incorporates the Parkes multibeam
  data in that the latter is the default maximum entropy model.
  Approximate heliocentric velocities in \kms\ are marked beside each
  significant component in the image.}
\label{f:atca-1616-moment}
\end{figure}

This object required even an even longer integration and a more compact
ATCA configuration to detect successfully with the ATCA.
J$1616-55$ is large, so a mosaic of 12 pointing positions each
separated by 15\arcmin, was required.

This object was imaged by making a dirty mosaic cube and beam cube following
the method of Sault et al. (1996). Natural weighting
was used. The cube was then
re-gridded to the velocity spacing of the multibeam data (which in turn was
re-gridded to the pixel size of the ATCA data) and deconvolved
using a maximum entropy task in {\sc miriad}. The multibeam data were used
as the `default image', thereby constraining the final deconvolved
image to be consistent with the ATCA data on small spatial scales, while
being consistent with the multibeam data on large spatial scales.
The resolution of the restored cube is 4\farcm5. Column-density and
velocity images were made
by applying a flux cutoff of 7 mJy beam$^{-1}$ to a smoothed version of the
cube. The final column-density image is shown in
Figure~\ref{f:atca-1616-moment}.

The ATCA observations show that the object appears to be resolved into two
bright components, at least four fainter components and, most
probably, a diffuse halo (Figure~\ref{f:atca-1616-moment}). The
brightest component lies in the southwest at 
RA $16^{\rm h}16^{\rm m}49^{\rm s}$, 
DEC $-55\arcdeg 44\arcmin 57\arcsec (\pm 1\arcmin)$
(J2000). A fainter `butterfly' feature lies about 25\arcmin\ to the
northeast at RA $16^{\rm h}19^{\rm m}05^{\rm s}$, DEC $-55\arcdeg
29\arcmin 08\arcsec$ (J2000). The velocities agree with the multibeam
observations: 430 km s$^{-1}$ in the southwest and 402 km s$^{-1}$ in
the northeast.  Because of the increased resolution, the peak column
density is higher in the ATCA image -- about $2.1\times 10^{19}$ atoms
cm$^{-2}$.

The morphology of this system is unusual. It may be a single, very low
column density galaxy where much of the disk has dropped below our
detection level. However, we suspect that, like J$1532-56$, it is
better explained by a tidal interaction between two smaller galaxies.
The fainter outlying features in Figure~\ref{f:atca-1616-moment} would
then represent tidal debris.  The combined \ion{H}{1} mass of the
whole system, if placed at 3.7 Mpc, is about a fifth of the mass of
that of the Small Magellanic Cloud. The projected separation between
the two main components is 27 kpc. The Galactic extinction in this
region is between $A_B=2.7$ and 4.4 mag (Table~\ref{t:newgal}). This
is low enough for a bright spiral such as J$1514-52$ to be seen, but
could easily mask a lower surface brightness system. We have inspected
Sky Survey plates of the region, and have taken MSSSO 40 inch CCD
frames in $V$, $R$ and H$\alpha$ (exposure times 5 -- 15 min) without
obtaining an optical identification. Thus we propose that J$1616-55$
is an interacting pair of binary LSB galaxies not unlike, though lower
in mass than, the \ion{H}{1} cloud discovered by Giovanelli \& Haynes
(1989).

The J$1616-55$ system appears to have no immediate extragalactic neighbors.
However, its heliocentric velocity (430 km s$^{-1}$, Table~1) is almost
identical to that of the Circinus galaxy (439 km s$^{-1}$, Jones et al. 1998).
Circinus lies at $\ell=311\fdg3, b=-3\fdg8$, about $18\arcdeg$ distant,
or a distance of only 1.2 Mpc if both objects are 3.7 Mpc distant.
Coincidentally, both galaxies lie at almost identical Galactic latitudes.
It is therefore a possibility that that Circinus is not an isolated galaxy, 
but has the J$1616-55$ system as a distant companion. In turn,
both galaxies in turn lie at the southern extremity of the Centaurus A
group which also has a mean distance of $\sim 3.5$ Mpc (Cot\'{e} et al. 1997).

\section{DISCUSSION}

The present sample is too small to predict the nature of
the galaxies that will be found in the Parkes multibeam surveys of the
Galactic Plane and the southern sky. It may, however, be significant that
tidal interactions are hypothesised for two of the three objects discussed
in this paper. Such systems are clearly under-represented in optical
surveys. The Magellanic Stream, for which no stars appear to be associated
(Wayte 1991), and the relatively recent discovery of the extent of the
tidal interaction in the M81 group (\S~\ref{s:1532-56}, Yun et al. 1994),
are good examples. The discovery of significant numbers of such systems may
put on a firmer footing the relationship between tidal interactions,
mergers and nuclear activity in nearby galaxies.

\begin{figure}[t]
\epsscale{0.8}
\plotone{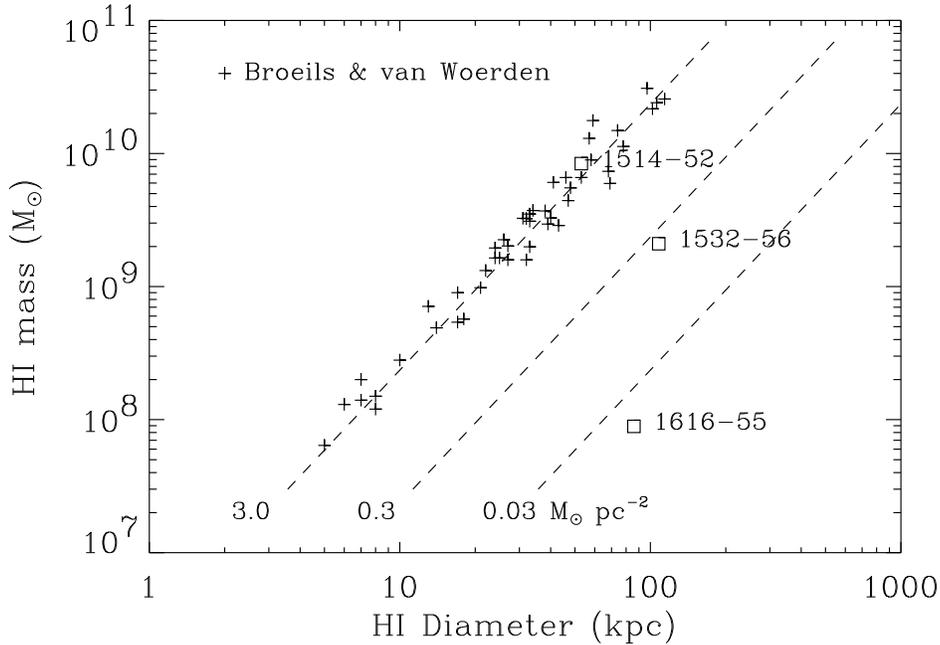}
\caption{\ion{H}{1} mass against \ion{H}{1} diameter for J$1514-52$, J$1532-56$
  and J$1616-55$ compared with the Broeils \& van Woerden (1994, BW)
  sample of large \ion{H}{1}-rich galaxies. Most of the BW sample and
  J$1514-52$ lie near the line representing an average surface density
  of 3 M$_{\odot}$ pc$^{-2}$ .  J$1532-56$ and J$1616-55$ have surface
  densities a factor of 13--200 smaller. Diameters for J$1514-52$ and
  the BW galaxies are measured at a surface density of 1 M$_{\odot}$
  pc$^{-2}$. The peak column density of J$1532-56$ and J$1616-55$
  lies beneath 1 M$_{\odot}$ pc$^{-2}$, so these were measured at 10\%
  of their peak surface densities, or $0.080$ and $0.017$ M$_{\odot}$
  pc$^{-2}$, respectively.}
\label{f:broeils}
\end{figure}

The same two objects (J$1532-56$ and J$1616-55$) are also significant
in having low column densities -- peaking at around $10^{19}$
cm$^{-2}$ at the multibeam resolution of $14'$ ($17'$ after gridding).
This corresponds to $<0.08$ M$_{\odot}$ pc$^{-2}$ projected \ion{H}{1}
column density, which is more than an order of magnitude below the
star-formation threshold proposed by Kennicutt (1989). The properties
of the new galaxies are compared in Figure~\ref{f:broeils} with the
optically selected sample of `normal' galaxies observed by Broeils \&
van Woerden (1994, hereafter BW) in a search for extended \ion{H}{1}
disks. The BW sample, which includes a wide range of morphological
types, \ion{H}{1} masses and luminosities, has a narrowly defined
range of mean \ion{H}{1} surface densities -- around 3 M$_{\odot}$
pc$^{-2}$.  J$1514-52$ has an \ion{H}{1} mass and an \ion{H}{1}
diameter (which has been measured at the same surface density of 1
M$_{\odot}$ pc$^{-2}$ -- Table~1) close to the high end of the BW
sample. It therefore has a similar mean (deprojected) surface density,
$\sim 3.8$ M$_{\odot}$ pc$^{-2}$.  In contrast, J$1532-56$ and
J$1616-55$ have mean surface densities of 0.23 and 0.015 M$_{\odot}$
pc$^{-2}$, respectively. The diameters of these objects are hard to
compare with their bright counterparts as their peak surface density
does not exceed the 1 M$_{\odot}$ pc$^{-2}$ level.  Similarly, the BW
data is too insensitive to measure diameters at a fainter level (i.e.\ 
BW would not have been able to detect J$1532-56$ and J$1616-55$ with
their observing parameters). Instead, we note that 1 M$_{\odot}$
pc$^{-2}$ is reasonably close to 10\% of the peak surface density for
galaxies in the BW sample, and use this criterion to define diameters
for J$1532-56$ and J$1616-55$. This level corresponds to contours at
$0.080$ and $0.017$ M$_{\odot}$ pc$^{-2}$, respectively.

What is the reason for the low \ion{H}{1} surface densities in
J$1532-56$ and J$1616-55$? Are these rare tidal systems? Or are these
common objects, akin to the optical low surface brightness galaxy
population (e.g.\ McGaugh 1996)? Morphologically, it appears that the
systems chosen may have had a strong tidal influence. However, it
should be noted that even the central \ion{H}{1} surface densities are
extremely low. For example, the peak (projected) surface density of
J$1616-55$ is a factor of 18 below the {\it mean} surface density
($\sim 3$ M$_{\odot}$ pc$^{-2}$) of the BW sample. The answer to
these questions awaits further multibeam observations. However, the
resolution of the Parkes telescope (14\arcmin) is such that the numbers
of low column-density galaxies which are close enough to be resolved
may be small.

\section{SUMMARY}

Three large galaxies have been found in a blind multibeam \ion{H}{1}
survey of the southern Galactic Plane. These galaxies are the first of
several thousand expected detections. One of them is a bright spiral
(or would be except for the several magnitudes of optical extinction) and has
been noted before by several authors. The remaining two are
new discoveries for which we have been unable to find optical/IR
counterparts, even with targeted observations. One of the new objects
lies precisely on the Galactic equator.

High-resolution (ATCA) \ion{H}{1} observations as well as optical CCD
observations have helped to reveal the nature of the three galaxies:

\begin{enumerate}

\item J$1514-52$ (SGC 1511.1-5249 = WKK 4748) is an inclined
  ($62^{\circ}$) disk galaxy with a flat rotation curve which extends
  to a radius of almost 60 kpc.  The \ion{H}{1} mass is $8.4\times
  10^{9}$ M$_{\odot}$ and the total mass is $5.7\times 10^{11}$
  M$_{\odot}$. This galaxy appears to have a moderately active nucleus
  $L_{\rm FIR}=1.0\times 10^{10}$ L$_{\odot}$.  This activity may be
  associated with having a nearby companion, ESO 223-G012.

\item J$1532-56$ appears as a very large (200 kpc) system in the
  multibeam images, although the ATCA diameter is 108 kpc.  Its
  complex velocity field suggests the presence of more than one
  galaxy, a view which seems to be supported by the ATCA images which
  show the system to have morphological similarities with other
  tidally interacting systems such as the M81 group. The Galactic
  equator cuts through the middle of this system, thus demonstrating
  the effectiveness of the blind \ion{H}{1} technique for finding
  galaxies even in the deepest obscuring layer of the Milky Way.

\item J$1616-55$ is the closest, lowest column density, and lowest mass
  of all the objects studied here. It appears to be a binary pair of
  LSB galaxies.  A high (optical) surface brightness galaxy would
  probably have been seen at this position, where the predicted
  obscuration is $A_B=2.7$ to 4.4 mag (although predictions such as
  these, based on IRAS data and Galactic \ion{H}{1} column densities,
  are extremely uncertain). This system is also interacting, and is
  possibly associated with the Circinus galaxy and/or the Centaurus A
  group.

\end{enumerate}

We speculate that interacting and low-column-density galaxies may be found
in significant numbers in  blind \ion{H}{1} surveys with the Parkes 
multibeam receiver and other systems.

\acknowledgments

The multibeam receiver was built through a collaborative venture
involving the ATNF (CSIRO), the University of Melbourne, Mount Stromlo
and Siding Spring Observatories, the University of Western Sydney, the
University of Wales at Cardiff, the Australian Research Council, and
Jodrell Bank. The superb design and construction efforts of Warwick
Wilson and his correlator group, and Mal Sinclair and his receiver
group are acknowledged.  The support of the aips++ group, on which all
of the basic data reduction software and some of the observing
software is based, is also gratefully acknowledged. Taisheng Ye
contributed much of the initial aips++ software. Much of the observing
and data reduction for the multibeam and ATCA observations was shared
with members of the HIPASS team. We gratefully acknowledge the HIPASS
team, especially Rachel Webster (PI), Alan Wright (operations), Ian
Stewart (operations) and David Barnes (reduction software). Rachel
Webster and David Barnes kindly obtained the CCD data presented in
this paper. We thank Jim Caswell, Jim Cohen, Richard Gooch and Andy Rivers for
additional contributions.


\end{document}